\documentclass{iopart}  
\usepackage{iopams}
\pdfoutput=1

\usepackage{graphicx}
\usepackage{color}

\definecolor{mygreen}{rgb}{0.0,0.4,0.25}

\newcommand{\rlj}[1]{{#1}}

\usepackage{hyperref}

\newcommand{\ee}{{\rm e}}
\newcommand{\beq}{\begin{equation}}
\newcommand{\eeq}{\end{equation}}

\begin{document}

\title[Accelerated relaxation in a kinetically constrained model with swaps]{Accelerated relaxation and suppressed dynamic heterogeneity in a kinetically constrained (East) model with swaps}

\author{Ricardo Guti\'errez}
\address{Complex Systems Group \& GISC, Universidad Rey Juan Carlos, 28933 M\'{o}stoles, Madrid, Spain}
\author{Juan P. Garrahan}
\address{School of Physics and Astronomy and Centre for the Mathematics and Theoretical Physics of Quantum Non-equilibrium Systems, University of Nottingham, Nottingham NG7 2RD, UK}
\author{Robert L. Jack}
\address{Department of Applied Mathematics and Theoretical Physics, University of Cambridge, Wilberforce Road, Cambridge CB3 0WA, United Kingdom}
\address{Department of Chemistry, University of Cambridge, Lensfield Road, Cambridge CB2 1EW, United Kingdom}

\begin{abstract}
We introduce a kinetically constrained spin model with a local softness parameter, such that spin flips can violate the kinetic constraint with an (annealed) site-dependent rate.  We show that adding MC swap moves to this model can dramatically accelerate structural relaxation.  We discuss the connection of this observation with the fact that swap moves are also able to accelerate relaxation in structural glasses.  We analyse the rates of relaxation in the model. We also show that the extent of dynamical heterogeneity is strongly suppressed by the swap moves.
\end{abstract}


\section{Introduction}

The glass transition is a dynamical phenomenon \cite{Binder2011,Berthier2011,Biroli2013}. In the supercooled regime, the structural relaxation time of a typical liquid behaves as 
\beq
\tau_\alpha = \tau_0 \exp\left( \frac{A(T)}{k_{\rm B} T} \right)
\label{equ:tau-alpha}
\eeq
where $\tau_0$ is a microscopic relaxation time, and $A(T)$ is a function (with units of energy) that increases on cooling.  It is natural to interpret $A(T)$ as a free energy barrier associated with structural relaxation, and a central aim of any theory of the glass transition is to explain the temperature-dependence of this quantity.  Some theories, including random first-order transition theory \cite{Berthier2011,Trachenko2011} and other mean-field theories of the glass transition \cite{Cavagna2009} propose that $A(T)$ can be determined on thermodynamic grounds -- the idea is that different amorphous states of the glass are separated by large free energy barriers, which must necessarily be crossed, in order for structural relaxation to take place.  The resulting picture is similar to classical nucleation theory, in which the free-energy barrier may be estimated from the surface tension and chemical potential difference, which are thermodynamic parameters.  Other theories of the glass transition, including dynamical facilitation theory \cite{Chandler2010}, observe that model systems with identical thermodynamic properties can (in general) have dramatically different relaxation times.  For example, if systems evolve by different dynamical rules then the energy barriers associated with dynamical relaxation can be very different.  Within such theories, the free-energy barrier $A(T)$ may be expected to depend strongly on the dynamical rules by which a glassy system evolves in time.

For atomistic systems, it is known that many properties of supercooled liquids are independent of microscopic details of the dynamics \cite{Binder2011,Berthier2011,Biroli2013}.  For example, systems where particles evolve by Newton's equations have very similar properties to those with Monte Carlo (MC) dynamics, and to overdamped Langevin dynamics~\cite{Berthier2007mc}.  This observation is consistent with thermodynamic pictures -- the interpretation is that the systems equilibrate quickly in local minima of the free energy, and the mechanisms by which they escape from these minima are controlled by thermodynamic barriers (which are independent of dynamics).  However, recent studies have shown that structural relaxation of some supercooled liquids can be accelerated by many orders of magnitude, by the simple addition of an extra dynamical process (MC move), in which particles of different sizes can swap their locations 
\cite{Grigera2001,Brumer2004,Fernandez2007,Gutierrez2015,Berthier2016}.  This has allowed the equilibration (by computer simulation) of supercooled liquid states with extremely large viscosities, comparable with experimental glasses \cite{Ninarello2017}.  This work has 
 enabled new computational studies of glassy states at low temperatures~\cite{Berthier2017entropy,Ozawa2018,Wang2019}. Several theoretical works have proposed explanations of the effects of swap dynamics~\cite{Wyart2017,Brito2018,Berthier2019-jcp,Ikeda2017,Szamel2018}.

This strong dependence of relaxation time on dynamics is unexpected within thermodynamic theories \cite{Wyart2017}, although explanations have been proposed, within an RFOT-like picture \cite{Ikeda2017,Berthier2019-jcp}.  In dynamical facilitation theory, it is natural to expect strong dependence of time scales on dynamics, but it is not clear why MC moves that swap particle sizes would have such a dramatic effect.  The predictions of facilitation theory are based on kinetically constrained models (KCMs) \cite{Ritort2003,Garrahan2011}, in which the degrees of freedom occupy the sites of a lattice, such as Ising spins in the case of facilitated spin models.  In such models, spin $i$ can only change its state if the neighbouring spins satisfy a constraint \cite{Ritort2003,Garrahan2011}. In this article, we consider softened KCMs \cite{Pan2005,Elmatad2010,Shokef2010}, in which the local structure of the liquid is accounted for by an additional variable on each site of the lattice, which we call the (local) softness.  If this quantity is large, there is a substantial probability that the system can relax locally, by violating the kinetic constraint.  We will show that even if the softness has a minimal effect on the natural dynamics of the model, introducing MC moves that swap the values of the local softness can dramatically accelerate relaxation.

The resulting models are simple ones, but we argue that the mechanism for acceleration by swaps may be general.  In a deeply supercooled liquid, most regions of the system are characterised by very large free energy barriers, but there are a few excitations that facilitate local motion \cite{Chandler2010,Keys2011}.  The swap mechanism accelerates dynamics because a region with a large local barrier can become soft via a swap move, which then enables relaxation.  In an atomistic system, one can imagine that a region might temporarily swap its particles for smaller ones; it can then relax, and then swap back the small particles for typically-sized ones.   That is, the free energy barrier for local relaxation can be lowered by a fluctuation in the local structure, which triggers relaxation.  This possibility is natural within our softened KCMs. 

The paper is organised as follows. In Section 2 we describe the soft KCM and the dynamical processes that will be considered, including different forms of local-softness swaps. In Section 3 we analyse theoretically the relaxation dynamics of the model. The inclusion of swaps in the dynamics is shown to be able to bring the system from a super-Arrhenius regime with a parabolic dependence of the relaxation time on the inverse temperature into an Arrhenius regime. In Section \ref{sec:numerics} we perform a numerical exploration of the model, which confirms the theoretical consequences derived in the previous section and allow us to investigate the effect of swaps on dynamic heterogeneity, as well as to compare different forms of local-softness swaps. In the concluding remarks, we discuss the implications of our results in the wider context of the physics of the glass transition.

\section{Model: Soft East KCM with swaps}

We describe several variants of the East model, with a local softness variables, and we explain how swaps are introduced as part of the set of rules that govern its dynamics. The local softness may be either a positive real number, or a binary variable. Both  models behave  similarly at low temperatures: the variant with binary softness is extensively used in the numerical explorations of Section \ref{sec:numerics}. 
\rlj{Different variants of the model require slightly different sets of parameters: these are summarised in Sec.~\ref{subsec:params},  below.  We emphasise that the main results of this work are robust across a broad range of parameters.}

\subsection{Soft East model}
\label{sec:global-soft}

The East model \cite{Jackle1991,Ritort2003,Garrahan2011} consists of $N$ binary spins $n_i\in\{0,1\}$ for $i=1,2,\dots,N$. For simplicity, we consider the one-dimensional case with periodic boundaries, so we identify spin $0$ with spin $N$ and spin $N+1$ with spin $1$. In contrast to the standard East model \cite{Jackle1991,Ritort2003,Garrahan2011},
 we consider a version where the kinetic constraint is ``soft'', cf.~\cite{Pan2005,Elmatad2010,Shokef2010,Elmatad2013,Gutierrez2016}. The model has two controlling parameters, $\epsilon$ and $c$, with $\epsilon\geq 0$ and $0<c<1$.  At site $i$ we define the softened  kinetic constraint  
\beq 
C_i = n_{i-1} + \epsilon 
\label{equ:constraint}
\eeq
If spin $i$ has $n_i=1$ then it flips to state $n_i=0$ with rate $C_i$; if $n_i=0$ then spin $i$ flips (to $n_i=1$)  with rate $C_i c/(1-c)$. The standard ``hard'' East model \cite{Jackle1991,Ritort2003,Garrahan2011} occurs for $\epsilon=0$, in which case spin $i$ is only updated when it is facilitated by its left neighbour.  For $\epsilon > 0$ spins can flip even when they are not facilitated: this is the sense in which $\epsilon$ softens the constraint.  The physical idea is that facilitated spins flip with a rate of order unity (``facilitated mechanism'') but all spins can also flip by an additional ``soft mechanism'', with a rate $\epsilon$ that is small at low temperatures.

In general, the parameters $c,\epsilon$ may both depend on temperature.  We take
\beq
c = \frac{1}{1+\ee^{J/T}}
\eeq
so that the system obeys detailed balance with respect to a Boltzmann distribution with energy $E=J\sum_i n_i$ (this fact is independent of $\epsilon$).  This Boltzmann distribution corresponds to independent spins with
\beq
\langle n_i \rangle = c \; .
\eeq
It is also natural to associate the softness $\epsilon$ with an energy barrier $U$ that may also depend on temperature, cf.\ \cite{Elmatad2010,Elmatad2013,Gutierrez2016}: 
\beq 
\epsilon=\ee^{-U(T)/T} \; .
\label{equ:epsU}
\eeq

The relaxation time of these models can be defined by considering the spin-spin correlation function $\langle n_i(t) n_i(0) \rangle - \langle n_i \rangle^2$, or the spectral gap of the generator.
For low temperatures (small $c$), the relaxation time of the hard ($\epsilon=0$) East model  is \cite{Sollich1999,Chleboun2013} 
\beq
\tau_{\rm East}  
\rlj{ \sim \exp\left( \frac{J^2}{2T^2 \ln2} + \frac{bJ}{T}\right) }
\label{equ:tau-east}
\eeq
\rlj{where $b$ is a constant (of order unity).}
The dominant contribution to the relaxation time is the super-Arrhenius term so the value of $b$ is relatively unimportant in the following.
\rlj{
We note at this point that this super-Arrhenius scaling appears because for an East model with (typical) excitation density $c$, the (free)-energy barrier for local relaxation scales as $T(\log c)^2$~\cite{Sollich1999,Chleboun2013}, so one may alternatively write (\ref{equ:tau-east}) as
\beq
\tau_{\rm East}  
\rlj{\sim \exp\left(\frac{(\log c)^2}{2\ln2} + \frac{b_0J}{T} \right) } \; ,
\label{equ:tau-east-c}
\eeq
where 
$b_0$ is a constant of order unity.
} 
The relaxation time of the softened East model  scales as
\beq
\tau_{\rm soft} 
  \sim {\rm min}\left(\tau_{\rm East} , 1/\epsilon \right) \; .
  \label{equ:tau-soft}
\eeq

Note also that in Sections \ref{sec:pers} and \ref{sec:numerics}, we also consider persistence times, which are typically of a similar order of magnitude to the relaxation time.  

\subsection{East model with local softness}
\label{sec:real}

We now introduce a fluctuating local softness. This is a qualitative departure from previous work on the soft East model, where the softness $\epsilon$ was spatially uniform and constant in time.  On each site $i$, we define an additional variable $X_i$, with units of energy.
We therefore replace the energy barrier $U$ in (\ref{equ:epsU}) by a site-dependent barrier which we write as
\beq
U_i = \left\{ \begin{array}{ll} B-X_i, & X_i<B \\ 0, &  \hbox{otherwise}\end{array} \right.
\label{equ:UB}
\eeq
where $B=B(T)$ is a parameter of the model, and $X_i$ is a (positive) energy associated with site $i$.  Physically, the energy barrier for the soft constraint is at most $B$; a large value of $X_i$ means that the local energy barrier is (relatively) small.  The model obeys detailed balance with respect to a Boltzmann distribution with 
\beq
\frac{E}{T} = \frac{J}{T} \sum_i n_i + \frac{1}{v(T)} \sum_i X_i
\label{equ:xn-boltz}
\eeq
where $v(T)$ is a (temperature-dependent) parameter.  This distribution is such that the spin variables and the local softness parameters are all independent with
\beq
\langle n_i \rangle = c = \frac{1}{1+\ee^{J/T}} \;, \qquad \langle X_i \rangle = v(T) \; .
\eeq
The probability density for $X_i$ is (for $X_i>0)$
\beq
p(X_i)=(1/v) \ee^{-X_i/v} \; .
\eeq

\subsubsection{No-swap dynamics.}

The dynamics of the model obeys detailed balance with respect to a Boltzmann distribution with energy (\ref{equ:xn-boltz}).  The spin $n_i$ has the dynamics of the soft-East model, with constraint function
\beq
C_i = n_{i-1} + \min(1,\ee^{-(B-X_i)/T})
\eeq
consistent with (\ref{equ:constraint},\ref{equ:UB}).  Note that flips of spin $i$ leave the associated $X_i$ unchanged.  

The dynamics of the $X$-variables is as follows: if $n_i=1$ then $X_i$ is updated (with rate $r_X$) to a new value drawn from the (equilibrium) exponential distribution $p(X)=(1/v) \ee^{-X/v}$.  If $n_i=0$ then $X_i$ does not update.  The rate $r_X$ is a parameter of the model.  The physical interpretation is that if there is an excitation on site $i$ ($n_i=1$) then the liquid structure at that site is exploring configuration space rapidly, so the local barrier $X_i$ is free to change.  On sites with $n_i=0$ then the local dynamics are very slow, and the $X_i$ cannot change, until such time as the $n_i$ variable flips.

\subsubsection{Swap dynamics.} 
\label{subsubsec:swaps}

For atomistic systems, the swap algorithm of \cite{Ninarello2017} means that the local softness in a given region of the system can change without requiring structural relaxation. For example, the size of the particles in that region might be reduced, which effectively reduces the softness barrier, and facilitates structural rearrangement.  To mimic this, we introduce an extra process by which the softness variables $X_i$ can be updated.  

We consider three possible choices for this extra process.  The first is called ``s-swaps'' (short for softness-swaps): with rate $N r_{\rm s}$, we choose two sites at random, and exchange their values of $X_i$.  The second is called ``s-updates'': with rate $N r_{\rm u}$ we choose a single site at random and update the local softness $X_i$ to a new value drawn from the (equilibrium) exponential distribution $p(X)=(1/v) \ee^{-X/v}$, similar to what was proposed in \cite{Brito2018}.  Note that this can happen on any site, independent of the value of $n_i$. The factors of $N$ in these rates are chosen so that the rates for individual spin flips are independent of system size. The third type of swap process is considered in Section \ref{sec:numerics}, namely, local swaps, which are s-swaps that only take place between neighbouring sites.  In this case, a random site is chosen with rate $N r_{\rm l}$, and its $X$-value is swapped with one of its neighbours (chosen at random).

\rlj{In a large system, s-swaps and s-updates lead to identical behaviour in the equilibrium state: this will be demonstrated explicitly in Fig.~\ref{fig:swap-upd}, below.   
To understand this, note first that they both obey detailed balance with respect to the Boltzmann distribution of (\ref{equ:xn-boltz}).   Second, note that if one considers separately the two spins participating in a swap, each one is updated to a value that is copied from the other spin, which will typically be far away (hence uncorrelated) and for which the softness distribution is the just same exponential used in an $s$-update.  In other words, from the point of view of a single spin $i$, an s-update is equivalent to swapping $X_i$ with a value chosen at random from a (fictitious) reservoir of softness values; alternatively one may swap the $X_i$ with a value from a (real) reservoir that is constituted by the other particles in the system (this is an s-swap).   If the reservoirs are large enough then correlation effects can be neglected and these two processes are equivalent.
Based on this equivalence, we refer to both s-swaps and s-updates as types of (non-local) swap moves.

Finally, note that an s-swap conserves the total softness $\sum_i X_i$ while an s-update does not.  However, the local (no-swap) dynamics of the $X$-variables already allows the total softness to fluctuate, so s-updates do not cause the breakdown of any (global) conservation law.  For this reason, the fact that s-swaps conserve the total softness is irrelevant, and one concludes that they are equivalent to s-updates. 
A recent study found a similar effect in an atomistic liquid~\cite{Berthier-new-swap}, where allowing particle diameters to be updated by local Hamiltonian dynamics leads to effects that are very similar to (non-local) swaps.  In that case the update rule does break a global conservation law (specifically, conservation of the number of particles with each value of the diameter), but an additional argument based on the equivalence of canonical and semi-grand ensembles establishes that updates have the same effect as swaps.}

\subsection{East model with binary softness}
\label{sec:binary}

In the following, we will show that the behaviour of the East model with local softness can be captured by an even simpler model, which we now describe.  For low temperatures, the effect of the local softness is dominated by sites with $X_i>B$.  The fraction of sites that have this property is easily verified to be $\ee^{-B/v}$.  

To exploit this fact, we replace each $X_i$ by a binary softness variable $s_i\in\{0,1\}$, such that sites with $s_i=1$ are soft.  
For consistency with the previous model, 
the constraint function is
\beq
C_i = n_{i-1} + s_i
\eeq
and the Boltzmann distribution for the $(n_i,s_i)$ has
\beq
\frac{E}{T} = \frac{J}{T} \sum_i n_i + \frac{B(T)}{v(T)} \sum_i s_i
\label{equ:sn-boltz}
\eeq
The dynamics of the $s$-variables are the same as those of the $X$-variables, except that where $X_i$ is updated with an exponentially-distributed random number, $s_i$ is updated to zero or $1$ with probability $(1+\ee^{-B/v})^{-1}$ and $(1+\ee^{B/v})^{-1}$ 
respectively.  The no-swap dynamics and the three varieties of swap dynamics are all generalised in this way.

Compared to the model with real-valued softness, 
this model introduces two simplifications: First, the binary variables $s_i$ enable efficient computer simulation of the model, 
\rlj{see the first paragraph of Sec.~\ref{sec:numerics} below.}
Second, the two parameters $B,v$ enter the binary model only through their ratio $B/v$.

\subsection{Summary of model parameters, and their physical interpretation}
\label{subsec:params}

Before continuing, we summarise the parameters \rlj{introduced so far.  We have chosen to maintain all rates as free parameters, so that our analysis is general. For this reason, our summary here also indicates which parameters are most important for controlling the qualitative behaviour of the model.}
\begin{itemize}
\item The parameter $J$ is the (free)-energy cost required to create an East-like excitation, $n_i=1$. This sets the fundamental energy scale in the model.  In the simulations below we set $J=1$. 

\item The parameter $B=B(T)$ is the maximal energy barrier associated with the soft constraint. The temperature-dependence of $B$ must be determined from physical arguments. In numerical simulations we typically take $B(T)\propto 1/T$ -- we  expect it to grow at low temperatures since the soft process is expected to be very slow in the glassy regime.

\item The parameter $v=v(T)$ determines the mean and the standard deviation of the softness parameter $X_i$. In general $v$ may depend on $T$ but in numerical simulations we take $v$ to be independent of temperature.

\item The parameter $r_X$ controls the rate with which the softness is updated on sites with $n_i=1$.    The time unit in the model is fixed by the facilitated process: the rates for facilitated spin flips are $1$ and $c$ (for flips into state with $n_i=0$ and $n_i=1$ respectively).    Since this is an excited site, one expects $r_X$ to be relatively fast.  In numerical simulations we take $r_X=\ee^{-J/T}$ which is much faster than the structural relaxation but small enough that adding the soft process to the simulation does not result in too much of an increase in the simulation efficiency.

\item The parameters $r_{\rm s},r_{\rm u}$ control the rates of (non-local) swap processes (Sec.~\ref{subsubsec:swaps}).  As for $r_X$, it is convenient (both numerically and analytically) to assume that these rates are small compared to $1$ but (very) large compared to the rate of structural relaxation.  In numerical simulations we mostly take $r_{\rm s}, r_{\rm u} \propto \ee^{-J/T}$, in which regime the results depend weakly on the specific choice of $r$.

\item The parameter $r_{\rm l}$ controls the rate of local swap processes. We take $r_{\rm l}$ of the same order as $r_{\rm s}$, that is $r_{\rm l} \propto \ee^{-J/T}$.  The results do depend significantly on this choice, as it determines the time scale of the diffusion of excitations, see Section~\ref{subsec:swap-type}.

\end{itemize}

\rlj{Note that the parameters $r_{\rm s},r_{\rm u},r_{\rm l}$ appear in different variants of the model.  For any given variant, only one of these parameters needs to be specified.}


\rlj{Anticipating the results that we derive below, we comment that the behaviour of the model depends strongly on $B$: if this parameter is too small then the soft relaxation process dominates the system and the system does not behave in a glassy way, regardless of whether swap moves are included.  The dependence on other parameters is much weaker.  In particular, our main result -- that s-swaps and s-updates dramatically accelerate relaxation -- does not require tuning of parameters.  This effect is robust as long as the relevant rates $r$ are not too small and $v$ is not too large. (These are the minimal constraints that are consistent with the underlying physics: If the $r$-rates are extremely small then the swap processes hardly ever happen so no acceleration is possible; also if $v$ is too large then soft process completely destroys the glassy behaviour, independent of whether swaps are present.)   These matters are discussed further in Sec.~\ref{sec:theory}.}
%


\section{Theory}
\label{sec:theory}

\subsection{No-swap dynamics}

\begin{figure}
\includegraphics[width=13cm]{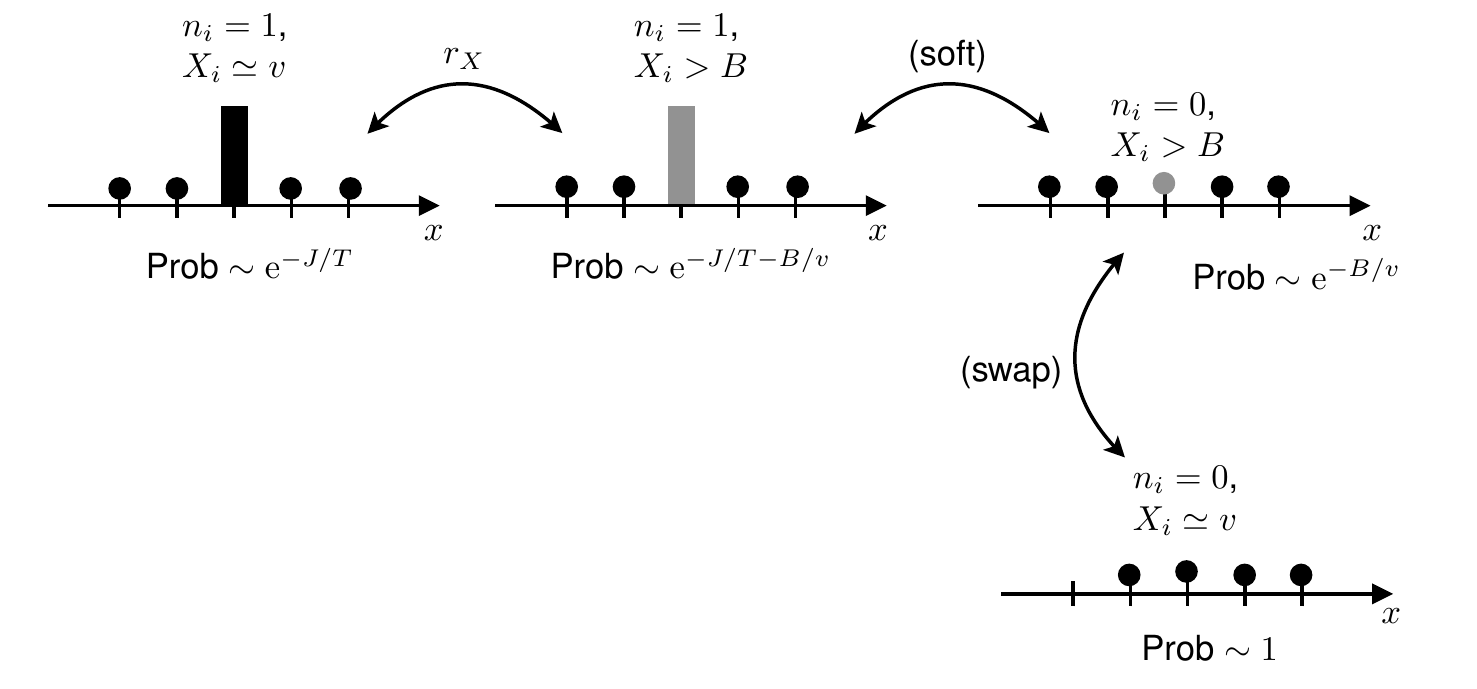}
\caption{Effects of the soft process and the swap process on East model configurations ($x$ indicates a spatial co-ordinate).  Vertical bars indicate spins with $n_i=1$ (as indicated) and circles indicate $n_i=0$.  Black colour indicates typical values of $X_i$, of order $v$.  Grey colour indicates anomalously soft sites, $X_i>B$.   For no-swap dynamics, the three configurations in the top row can interconvert by the $r_X$ process (which refreshes $X_i$ with rate $r_X$) and the soft process (which leads to ``fast'' flips if $X_i>B$).  The probabilities indicate the equilibrium probabilities of these states, assuming that $\ee^{-J/T},\ee^{-B/v}\ll 1$.  If this interconversion is fast compared to structural relaxation, one may interpret the three states as different manifestations of a single ``effective excitation'', which spends a fraction $f$ of its time in the top right configuration.  If swapping is enabled, the ``vertical'' process shown at right is also possible: this allows removal of the ``effective excitation'' by a purely local process and dramatically accelerates the relaxation.  [For no-swap dynamics, the effective excitation can only be destroyed by a co-operative (facilitated) process, or by a rare event where a spin with $X_i\simeq v$ flips spontaneously by the soft process.  If (\ref{equ:accel-condition}) is satisfied, then such events are much rarer than the ones shown in this figure, and the swap move strongly accelerates the dynamics.]}
\label{fig:swap-mech}
\end{figure}

We first consider the behaviour of the East model with local softness, in the absence of swaps.   The typical barrier for the soft constraint is $B-v$.  We assume that this barrier is large enough that the no-swap dynamics are controlled by the facilitated (East) relaxation process.  This requires (as a necessary condition) that the \emph{typical} barrier for the soft process is large, specifically
\beq
B(T)-v(T) \gtrsim T \log \tau_{\rm East} \;   .
\label{equ:soft-typ}
\eeq
For this condition to hold at low temperatures, it is necessary that $B$ must increase as $T$ is reduced.  In this regime, the barrier for the soft process is large, and one expects the behaviour to be dominated by the facilitated process.  However, the soft process may still have important effects, which come from (non-typical) sites where $X\gtrsim B$, as we now discuss.   

Any site with $X_i>B$ that has $n_i=0$ will flip with a rate of order unity into a state with $n_i=1$.  At that point, the value of $X_i$ will be updated, so it will (most likely) revert to a typical value $X_i\simeq v$.  This means that sites with $X_i>B$ quickly convert into excited sites (with $n_i=1$).  The reverse process is also possible: an excitation ($n_i=1$) can convert to an unexcited but softened state ($n_i=0$ but $X_i>B$).  These processes are illustrated in the top row of Fig.~\ref{fig:swap-mech}.  The overall effect of this interconversion is that long-lived ``effective excitations'' (or ``superspins''~\cite{Sollich1999}) spend some of their time in the unexcited but softened state. 
(This argument requires that $1/r_X$ is much smaller than the lifetime of a typical effective excitation; that is, interconversion between the two states is faster than structural relaxation.  We take $r_X\sim\ee^{-J/T}$ but structural relaxation at low temperatures is much slower,  
so this is satisfied in practice.) 

\rlj{Now define $\theta_i$ to be equal to unity if site $i$ is soft ($X_i>B)$ and zero otherwise, so that $\langle\theta_i\rangle = \ee^{-B/v}$.  Then the density of the effective excitations is the fraction of sites with either $n_i=1$ or $\theta_i=1$ (or both), 
which is
\beq
c_{\rm eff} = \left\langle n_i + \theta_i - n_i \theta_i  \right\rangle\; .
\label{equ:c-eff}
\eeq
Since both $\ee^{-(B/v)}$ and $\ee^{-(J/T)}$ are small then the $\langle n_i \theta_i\rangle$ is negligible and
we obtain $c_{\rm eff} \approx \ee^{-B/v} + \ee^{-J/T}$.  
Hence
one sees that the fraction of time spent by a effective excitation in the softened state (with $\theta_i=1$) is
$$
f = \frac{ \langle \theta_i \rangle}{ c_{\rm eff} } \approx \frac{\ee^{-B/v}}{\ee^{-B/v} + \ee^{-J/T}} \; .
$$
Using (\ref{equ:c-eff}) to write $\langle\theta_i\rangle \approx c_{\rm eff}-c$, one obtains
$$
c_{\rm eff} \approx \frac{c}{1-f} \approx c \left( 1 + \ee^{(J/T)-(B/v)} \right)
$$
where the approximate equalities are all valid up to terms of order $\ee^{-(B/v)-(J/T)}$.
For the model with binary softness, the behaviour is the same on replacing $\theta_i$ with $s_i$.}

We see that for the system to behave similarly to an East model (with $c_{\rm eff}\approx c$), we require at low temperatures that
\beq
\ee^{-B/v} \ll \ee^{-J/T} ,
\label{equ:soft-rare}
\eeq
which also implies $f\approx \ee^{(J/T)-(B/v)} \ll 1$.
Hence, while (\ref{equ:soft-typ}) is necessary for the system to behave similarly to the original (hard) East model, it is not sufficient, since (\ref{equ:soft-rare}) is also required.  

Given (\ref{equ:soft-typ},\ref{equ:soft-rare}) we expect [by analogy with (\ref{equ:tau-east-c})] that the relaxation time of the models with local softness is
\beq
\tau_{{\rm noswap}}  \sim \exp\left(\frac{(\log c_{\rm eff})^2}{2\ln 2} + \frac{b_0J}{T} \right)
\label{equ:tau-noswap}
\eeq
\rlj{This corresponds to a speedup of the East dynamics, due to the softening of the constraint.  For the parameters considered here we will have $c_{\rm eff}/c\approx 1$, so this effect is weak.  That is, we focus in the following on a dynamical speedup effect that is due to swaps (or updates) of the softness; it is not a simple consequence of the softened constraint.  We note however that the very strong dependence of $\tau_{{\rm noswap}} $ on $c_{\rm eff}$ means that this weak speedup is still observable even when the differences between $c_{\rm eff}$ and $c$ are small, see Sec.~\ref{sec:numerics}
and in particular Fig.~\ref{fig:auto}.}


\subsection{Effect of (non-local) Swap dynamics}

\subsubsection{Relaxation time}
When swap dynamics are included, a new process becomes important.  In this case, a spin with $n_i=0$ and $X_i>B$ can still interconvert with an excitation by the mechanism described above.  However, it is also possible that this spin can swap its $X$-value with another spin for which $X$ has a typical value (of order $v$).  This provides a mechanism by which an effective excitation can be converted to an unexcited state with $(n_i,X_i)=(0,v)$.  This is shown as the rightmost (vertical) process in Fig.~\ref{fig:swap-mech}.  The net effect of the chain of processes in Fig.~\ref{fig:swap-mech} is the same as 
 that of the original soft process of Sec.~\ref{sec:global-soft}: a site with $n_i=1$ can convert to $n_i=0$, without ever becoming facilitated, and with typical values of the local softness in both the initial and final states.
\rlj{For the model with $s$-update moves, the rate for spontaneous destruction of an excitation can be estimated as (approximately)
\beq
\epsilon_{\rm eff} =  r_X \ee^{-B/v} \cdot \frac{1}{1+r_X} \cdot \frac{r_{\rm u}}{r_{\rm u} + \ee^{-J/T}} \; .
\label{equ:spont}
\eeq
}%
To see this, we consider the sequence of processes obtained by reading from left to right in Fig.~\ref{fig:swap-mech}):
the first factor in (\ref{equ:spont}) is the rate for the initial transition in Fig.~\ref{fig:swap-mech}, the second factor is the probability that the excitation is destroyed (second step) before the system reverts back to the first state.  Similarly, the third factor is the probability that the system makes the final (swap) step before the second step gets spontaneously reversed.   \rlj{For the model with $s$-swaps, a similar formula holds, with $r_{\rm u}$ replaced by $r_{\rm s}$.
By analogy with (\ref{equ:tau-soft}), one infers that the relaxation time in a system with (nonlocal) swaps is
\beq
\tau_{{\rm swap}}  \sim \min\left(  1/\epsilon_{\rm eff} , \tau_{\rm noswap} \right) \; .
\label{equ:tau-swap-general}
\eeq
}

The estimate in (\ref{equ:spont}) is not expected to give quantitative predictions but it should capture the scaling of $\epsilon_{\rm eff}$.  From Sec.~\ref{subsec:params} we have $r_X\ll 1$ at low temperatures, while $r_{\rm u}$ and $\ee^{-J/T}$ are of a similar order.
The result is that the ``effective softness parameter'' is
\beq
\epsilon_{\rm eff} \approx r_X \ee^{-B/v} .
\label{equ:eps-eff-rlarge}
\eeq
This is valid for $r_{\rm u}\gtrsim \ee^{-J/T}$ and $r_X\ll1$.  If on the other hand $r_{\rm u} \ll \ee^{-J/T}$ then \
\beq
\epsilon_{\rm eff} \approx r_X r_u \ee^{-B/v+J/T}. 
\label{equ:eps-eff-rsmall}
\eeq
In the former case (which includes $r_{\rm u}\approx\ee^{-J/T}$), we predict that
\beq
\tau_{{\rm swap}}  \sim \min\left(  r_X^{-1} \ee^{B/v} , \tau_{\rm noswap} \right) \; .
\label{equ:tau-swap}
\eeq
%

\subsubsection{Implications}
Suppose that (\ref{equ:soft-typ},\ref{equ:soft-rare}) hold, so that the dynamics without swaps depends very weakly on the fact that the model is soft; and assume also that
\beq
\rlj{(\tau_{\rm noswap})^{-1} \ll \epsilon_{\rm eff} \ll 1} \; .
\label{equ:accel-condition}
\eeq
Then (\ref{equ:tau-swap-general}) shows that the swap dynamics will be accelerated dramatically, compared to the no-swap dynamics.   
\rlj{That is
\beq
1 \ll \tau_{\rm swap} \ll \tau_{\rm noswap} \; 
\label{equ:accel-result}
\eeq
This is the key theoretical result of this paper, which we verify in Sec.~\ref{sec:numerics} by numerical simulations.
It compares two models which have the same (softened) kinetic constraint, the only difference is whether there are swap moves that update the local softness.

As anticipated in Sec.~\ref{subsec:params}, Equ.~(\ref{equ:accel-result}) requires some assumptions on the rates in the model, particularly (\ref{equ:soft-typ},\ref{equ:soft-rare}) which require that the constraint is not so soft that all glassy behaviour is destroyed, independent of swaps.  We also require (\ref{equ:accel-condition}) which is 
an assumption on the $r$-rates (that is, $r_{\rm s},r_{\rm u},r_X$).
 From (\ref{equ:soft-rare}) we have $\ee^{-B/v}\ll 1$; together with (\ref{equ:spont}) this means that (\ref{equ:accel-condition}) fails only if one of the 
 $r$-rates is extremely small, of the order of $(\tau_{\rm noswap})^{-1}$.  Recall that $(\tau_{\rm noswap})^{-1}$ is the rate for a highly-collective relaxation process and has a super-Arrhenius dependence on temperature.  On the other hand, there is no physical justification for choosing very small values for the $r$-rates, which are microscopic parameters, and should therefore be much larger than $(\tau_{\rm noswap})^{-1}$ at low temperatures.  Hence (\ref{equ:accel-condition}) holds very generally, and 
 (\ref{equ:accel-result}) is a robust result, independent of specific choices of the rates $r$.}

As a specific case, we take (consistent with Sec.~\ref{subsec:params}) that $B(T)\sim aJ^2/T$ with $a>1$ and $v$ is independent of temperature, then (\ref{equ:soft-typ},\ref{equ:soft-rare},\ref{equ:accel-condition}) are all satisfied, and we find $B/v \propto 1/T$.  
In this case the swap dynamics will have an Arrhenius temperature dependence, while the no-swap dynamics would be super-Arrhenius.  This case is discussed in more detail \rlj{in Sec.~\ref{sec:num-accel}, below.}

More generally, it is useful to consider the mechanism of acceleration by swaps.  We assume that with some small probability, the system has a reduced energy barrier for local relaxation.  Without swap dynamics, these soft regions enhance the rates for local motion, but their effects are confined to a particular region.  The effect of the (non-local) swaps is that locally-inactive regions can be (temporarily) activated by ``importing'' softness from other regions of the system.  This facilitates local relaxation, after which the softness can be exported away to some other inactive region.  In (polydisperse) atomistic systems, one might imagine that softness is imported by swapping some large particles with smaller particles from elsewhere, which increases the local free volume and facilitates relaxation.  After this relaxation has taken place, the small particles can be exported away again, and large particles can be re-imported.

\rlj{
We emphasise that the swap moves do not change the excitation variables $n_i$ directly.  Rather, the (non-local) swap moves change the softness, which then triggers (local) relaxation.  Note also that this mechanism is relatively insensitive to the precise route by which softness is imported -- this might be a non-local swap~\cite{Ninarello2017} or a deterministic local change in particle diameter~\cite{Berthier-new-swap}, or any other mechanism for accessing local configurations where the barrier for structural relaxation is reduced.}

It is also noteworthy that for the models considered here, the swap mechanism does not have any collective character.  Just like the soft process in the model of Sec.~\ref{sec:global-soft}, the swap-mediated relaxation can destroy or create excitations, independently of their environment.  This means that (for example) the relaxation will be less dynamically heterogeneous in the presence of the swaps: see also Fig.~\ref{fig:chi4} below.  
\rlj{However, the microscopic mechanism for the \emph{soft process} in liquids should have some collective character.  The nature of this process is not explicit within the model: it presumably determines parameters such as the size of the large energy barrier $B$ and the temperature-dependence of $B/v$.  It is not possible to explore these effects within this model} -- one would (presumably) have to consider the liquid structure and the way that the particles are packed in space.

\subsection{Persistence function}
\label{sec:pers}

So far we have concentrated on the relaxation time, which we defined as the time scale associated with the decay of the the spin correlation function $\langle n_i(t) n_i(0) \rangle - \langle n_i \rangle^2$.  For numerical work it is also useful to consider the persistence function.  
This is defined in terms of the local {\em persistence field}: the local variable $p_i(t)$ takes value $p_i(t)=1$ if $n_i$ has not flipped from its initial state at time $0$ up to time $t$, otherwise $p_i(t)=0$.  The persistence function is 
\beq
P(t) = \langle p_i(t) \rangle
\eeq
where the right hand side is independent of $i$, by translational invariance. 
\rlj{The behaviour of this function has been studied extensively in kinetically constrained models~\cite{Garrahan2011,Garrahan2002,Jung2005,Pan2005-het,Teomy2015}.}
The persistence time $\tau_{\rm p}$ corresponds to the typical timescale for the decay of $P(t)$. We define it through a threshold
\beq
P(\tau_{\rm p}) = 10^{-2}.
\label{p100}
\eeq
(Other definitions, such as through the time integral of $P$, give results that scale similarly with the parameters of the model.)  Our numerical work mostly focusses on the persistence function instead of the autocorrelation, as it is numerically easier to estimate and contains similar information.  For (hard) kinetically constrained models, the persistence and correlation times are similar.  However, it is important to note that when relaxation occurs by the soft process, the persistence time is larger than the relaxation times that we have considered so far.

In particular, if the kinetic constraint is completely irrelevant (that is, spins are non-interacting and flip only by the soft process) then it is easy to show that the persistence time is longer than the correlation time by a factor of $\ee^{J/T}$.  The physical reason is that the autocorrelation time is equal to the typical time taken for a typical spin with $n_i=1$ to decay to $n_i=0$, which is the inverse of the rate for the slow process, $\epsilon^{-1}$.  After a few multiples of this time, the state of the system has decorrelated.  However, spins with $n_i=0$ are much more numerous at low temperatures so they dominate the persistence time, which is comparable to the time taken for a typical spin with $n_i=0$ to flip to $n_i=1$.  This time is of order $\epsilon^{-1}\ee^{J/T}$.

In the softened East models considered here, the decoupling between persistence and autocorrelation is less strong than this extreme case, where spins are completely independent.  We will see in the following that the persistence time is significantly larger than the correlation time.  It is also notable that if relaxation takes place by the soft mechanism, one expects exponential relaxation of both the correlation and the persistence function $P(t)$.  

\subsection{Effect of local Swap dynamics}
\label{sec:local-swap-theory}

We briefly discuss effects of local swap dynamics.  In this case Fig.~\ref{fig:swap-mech} is modified only in the final (vertical) process, in which the anomalously soft (grey) site would not disappear.  Instead, it would hop to an adjacent site.  For $r_{\rm l} \ll \ee^{-J/T}$, the excitation typically converts back to $n_i=1$ after hopping, by reversing the processes in the top row of Fig.~\ref{fig:swap-mech}.  This leads to excitation diffusion with a hop rate of order $r_Xr_{\rm l}\ee^{-B/v+J/T}$ [similar to (\ref{equ:eps-eff-rsmall})] and a hop size of one lattice spacing, so the resulting diffusion constant is 
\beq
D_{\rm eff} \sim r_X r_{\rm l} \ee^{-B/v+J/T}.
\label{equ:Deff}
\eeq
On the other hand, for $r_{\rm l} \gtrsim \ee^{-J/T}$, the excitation may hop several times before converting back to $n_i=1$ (this conversion happens with rate $\ee^{-J/T}$).  In this case we expect that excitations hop with a rate of order $r_X \ee^{-B/v}$ [by a similar argument to (\ref{equ:eps-eff-rlarge})]; the typical size of a single hop is of order $\sqrt{r_{\rm l} \ee^{J/T}}$.  This leads to the same diffusion constant as (\ref{equ:Deff}).
Note that if the hop size is larger than the typical excitation spacing $\ee^{J/T}$,
we expect the behaviour of the model to resemble that for non-local s-swaps (with $r_{\rm s}=r_{\rm l}$). 
This requires a very large rate $r_{\rm l} \gtrsim \ee^{J/T}$. However, in higher dimensions ($d\geq2$), the typical excitation spacing is much smaller, scaling as $\ee^{J/(Td)}$.  We expect the other arguments of this section to depend weakly on dimension: in this case local and non-local swaps should lead to similar behaviour as long as $r_{\rm l}\gtrsim \ee^{J((2/d)-1)/T}$.  This constraint is much weaker in higher dimensions.
 
We focus on $r_{\rm l} \propto \ee^{-J/T}$ consistent with Sec~\ref{subsec:params}, which leads to short-ranged hopping of the excitations.  The resulting dynamics resembles a Fredrickson-Andersen (FA) model~\cite{Garrahan2011}, with excitations that diffuse.  A notable signature of this behaviour is the persistence function -- one expects that for $P(t)=O(1)$ then 
\beq
P(t) \sim \exp(-\sqrt{D_{\rm eff} c^2 t})
\label{equ:P-fa}
\eeq
 where $D_{\rm eff}$ is the excitation diffusion constant and $c$ their separation~\cite{Garrahan2002}. For very large times, $P(t)$ crosses over to an exponential decay~\cite{Garrahan2011}, but $P$ itself is very small in this limit.
The prediction of this analysis is that persistence time decays significantly slower than the relaxation with local swaps: one may estimate
\beq
\tau_{\rm p}^{\rm loc-swap} \sim  \min\left(  \frac{1}{r_{\rm l}r_{X}} \ee^{B/v+J/T} , \tau_{\rm noswap} \right)
\label{equ:pers-loc-swap}
\eeq
This prediction is rather simplistic because it assumes that the diffusive (FA) process dominates the relaxation. 
In practice, the original facilitated process of the East model is also expected to play a role, so the relaxation is a mixture of FA and East dynamics.  Compared to the case where the FA dynamics dominates, the East dynamics acts as an additional relaxation channel, so one expects relaxation to be somewhat faster than that predicted by (\ref{equ:pers-loc-swap}).

%

\begin{figure}
\includegraphics[width=14cm]{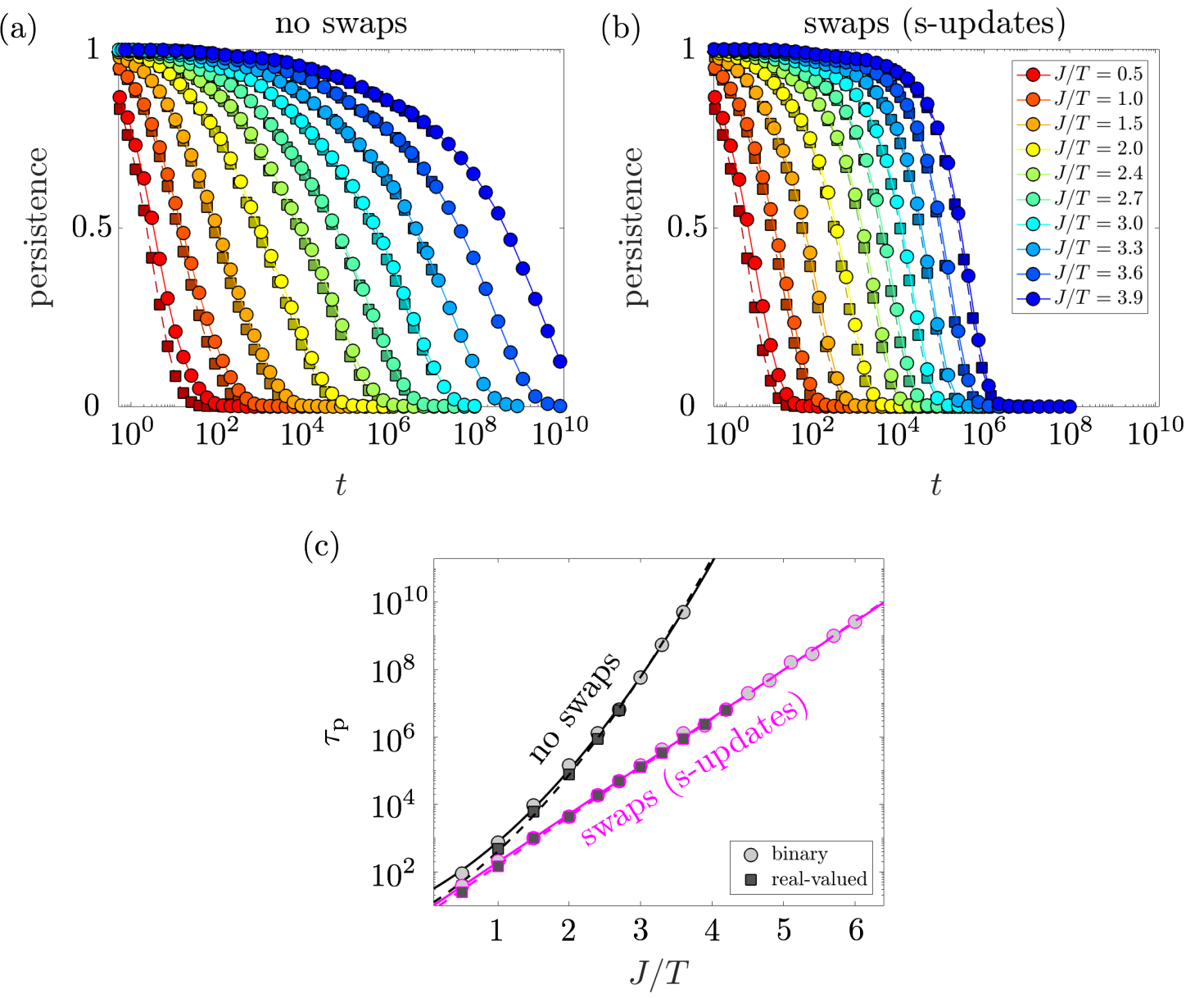}
\caption{Persistence function for the East model with real-valued local softness (squares, dashed lines) and binary softness (circles, continuous lines).  We take $B/v = 2J/T$ (in the case of the real-valued model, $B=10 J/T$ and $v=5$), for various values of $J/T$ (see legend), in a system of $N=512$ spins. 
(a)~Dynamics without swaps (b)~Dynamics including swaps (s-updates) 
(c)~Dependence of the persistence time $\tau_{\rm p}$ on temperature for both variants of the model, with fits $\tau_{\rm p}=\tau_0 \exp\left[ (bJ/T) + J^2/(2T^2\ln 2)\right]$ for dynamics without swaps [see (\ref{equ:tau-noswap})] and $\tau_{\rm p}=\tau_0 \exp\left( bJ/T \right)$ with swaps [see (\ref{equ:tau-swap})] 
(the fitting parameters are $b,\tau_0$).  The fits are excellent for $J/T>1$.
}
\label{fig:pers}
\end{figure}

\section{Numerical results}
\label{sec:numerics}

We have performed numerical simulations of the East models with local softness, \rlj{using variants of the Bortz-Kalos-Lebowitz (BKL) algorithm, which is also known as continuous-time Monte Carlo~\cite{Bortz1975}.} For the simpler model with binary softness $s_i$, where the set of rates of local moves is finite, the standard rejection-free BKL method is used. For the version of the model with real-valued softness $X_i$, and with energy barrier as in Eq.~(\ref{equ:UB}), where the set of possible transitions is not finite, we adapt the BKL algorithm: the different types of moves ($n_i=0 \rightarrow n_i = 1$, $n_i=1 \rightarrow n_i = 0$, and updates of the local softness $X_i$) are proposed with rates that depend on $n_i$ and $n_{i-1}$ but not on $X_i$. In order that MC moves occur with the correct rates, the moves that are proposed in this way are then accepted with a probability that accounts for the dependence of the rate on $X_i$.

The remainder of this section presents numerical results, with a special emphasis on the impact of swaps on the relaxation dynamics at low temperatures. First, we focus on the temperature dependence of relaxation and persistence times for dynamics with and without swaps. We then study the implications of including swaps for the dynamic heterogeneity of the relaxation process. To conclude, we compare different types of swap processes. Throughout the section, numerical results are related to the theoretical arguments provided in the previous sections.  A more general discussion of these results (including comparisons with atomistic systems) and their implications is given in Section~\ref{sec:discuss}.

\subsection{Acceleration due to swaps (s-updates)}
\label{sec:num-accel}

To illustrate the main behaviour, we consider the time dependence of the persistence function for a range of temperatures in the East model with a real-valued local softness (Sec~\ref{sec:real}) and with binary local softness (Sec~\ref{sec:binary}).  We consider dynamics without swaps, and also a model with s-updates as swap processes; results for s-swaps and local swaps are given in Sec.~\ref{subsec:swap-type}, below.  We take $B/v=2J/T$: this ratio is itself a parameter of the model with binary softness; in the model with real-valued softness, we take $B=10 J/T$ and $v=5$.  The swap rate is $r_{\rm u} = \ee^{-J/T}/4$ in both cases, consistent with Sec. \ref{subsec:params}.  This choice is also consistent with (\ref{equ:soft-typ},\ref{equ:soft-rare},\ref{equ:accel-condition}), so we predict that the softening will have a weak effect on the no-swap dynamics, but the swap dynamics will lead to significant acceleration.

Figure~\ref{fig:pers} compares the relaxation with and without swaps. Figure \ref{fig:pers}(a) shows the persistence as a function of time for various inverse temperatures (in units of $J$) in the absence of swaps.  We show data for the model with real-valued softness, and for the version with binary softness. The two variants of the model behave almost identically at low temperatures (but the binary-softness can be simulated to much longer times due to its simplicity). Figure \ref{fig:pers}(b) shows the persistence for the same conditions but now including swap dynamics (corresponding to s-updates). As predicted, one sees a dramatic acceleration of relaxation by the swap dynamics, consistent with Eqs.~(\ref{equ:tau-noswap},\ref{equ:tau-swap}). Figure \ref{fig:pers}(c) shows the corresponding persistence times $\tau_{\rm p}$ as functions of $J/T$ for both variants of the model. The data are compatible with a super-Arrhenius scaling for the dynamics without swaps [see (\ref{equ:tau-noswap})] and an Arrhenius dependence in the presence of swaps [see (\ref{equ:tau-swap})]. As in the case of atomistic models \cite{Ninarello2017} the inclusion of swaps allows to significantly increase the range of low temperatures accessible to the numerics. 
\rlj{We emphasise that the results of Fig.~\ref{fig:pers} are all for models with softened constraints -- the speedup is not a simple consequence of the softening, it comes instead from the s-updates, which trigger local relaxation of the excitations.}

\begin{figure}
\includegraphics[width=14cm]{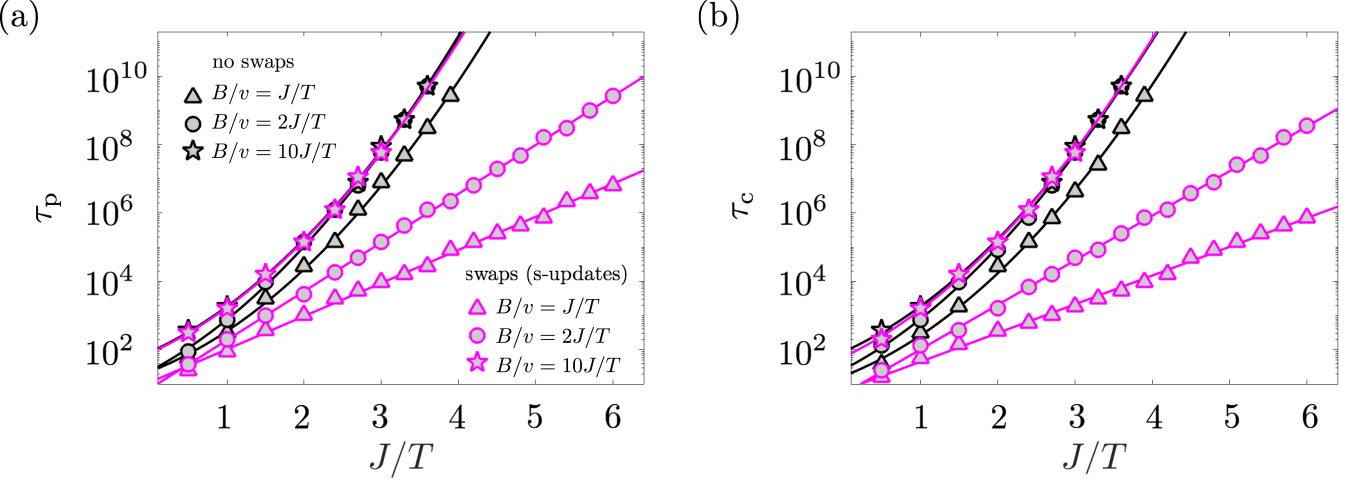}
\caption{Persistence time (a) and autocorrelation time (b) as functions of $J/T$, for the East model with binary local softness. 
The softness parameter $B/v$ is varied, as shown in the legend.  The system size is $N=512$ spins.  
The rate for the soft process is smallest when $B/v=10J/T$ -- in this case, the dynamics is almost unaffected by the soft process, both with and without swaps.
By contrast, the soft process is relatively fast when $B/v=J/T$ -- this has a strong effect on the dynamics with swaps, and a weak (but significant) effect on the no-swap dynamics.
The solid lines are fits either to $\tau=\tau_0 \exp\left[ (bJ/T) + J^2/(2T^2\ln 2)\right]$, in the case of the dynamics without swaps or dynamics with swaps for $B/v=10J/T$, or $\tau=\tau_0 \exp\left( bJ/T \right)$, in the case of the dynamics with swaps for $B/v=J/T$ and $2J/T$.  See the discussion in the main text.
}
\label{fig:auto}
\end{figure}

Since the models with real-valued and binary softness behave the same, we focus on the model with binary softness in the following.   Fig.~\ref{fig:auto}(a) shows how the persistence time depends on the choice of $B/v$: we take $B/v=yJ/T$ with $y=1,2,10$. 
As predicted by (\ref{equ:tau-swap}), for large $y$ (in this case $y=10$) the swap mechanism is negligible and the systems with and without swaps behave the same.  For $y=2$ one observes a dramatic acceleration by the swaps, as in Fig.~\ref{fig:pers}.  For $y=1$, the condition (\ref{equ:soft-rare}) is breaking down, even though (\ref{equ:soft-typ}) is still satisfied at low temperatures.  As predicted by (\ref{equ:tau-noswap}), this means that the soft process leads to a reduction in the relaxation time, even in the absence of swaps (because $c_{\rm eff}>c$).  Nevertheless, the swaps still lead to a significant acceleration.

Figure~\ref{fig:auto}(b) shows that analogous results are obtained by considering the correlation time $\tau_{\rm c}$ extracted from the autocorrelation ($\langle n_i(t) n_i(0) \rangle - \langle n_i \rangle^2$) with the same threshold as in (\ref{p100}).   From the fits in this Figure, it is useful to consider swap dynamics with $B/v=J/T$ and $B/v=2J/T$.  In this case the fits are $\tau_{\rm c}=\tau_0 \exp\left( bJ/T \right)$ with $b=(1.9,3.0)$ for $B/v=(J/T,2J/T)$ respectively.  Recalling that $r_X\sim \ee^{-J/T}$, Equ.~(\ref{equ:tau-swap}) predicts $b=(2,3)$, consistent with these data.  In these cases one also finds that the persistence time is significantly larger than the correlation time, consistent with the arguments of Sec.~\ref{sec:pers}.  Fits for the persistence times $\tau_{\rm p}$ yield $b=(2.2,3.3)$.  If the relaxation was entirely dominated by the soft mechanism one would expect $b=(3,4)$ -- the observed values of $b$ are intermediate between these values and the exponents of the correlation time, consistent with Sec.~\ref{sec:pers}.  For the cases where the facilitated process is dominating, we observe good fits to $\tau=\tau_0 \exp\left[ (bJ/T) + J^2/(2T^2\ln 2)\right]$, consistent with (\ref{equ:tau-noswap}).

\begin{figure}
\includegraphics[width=14cm]{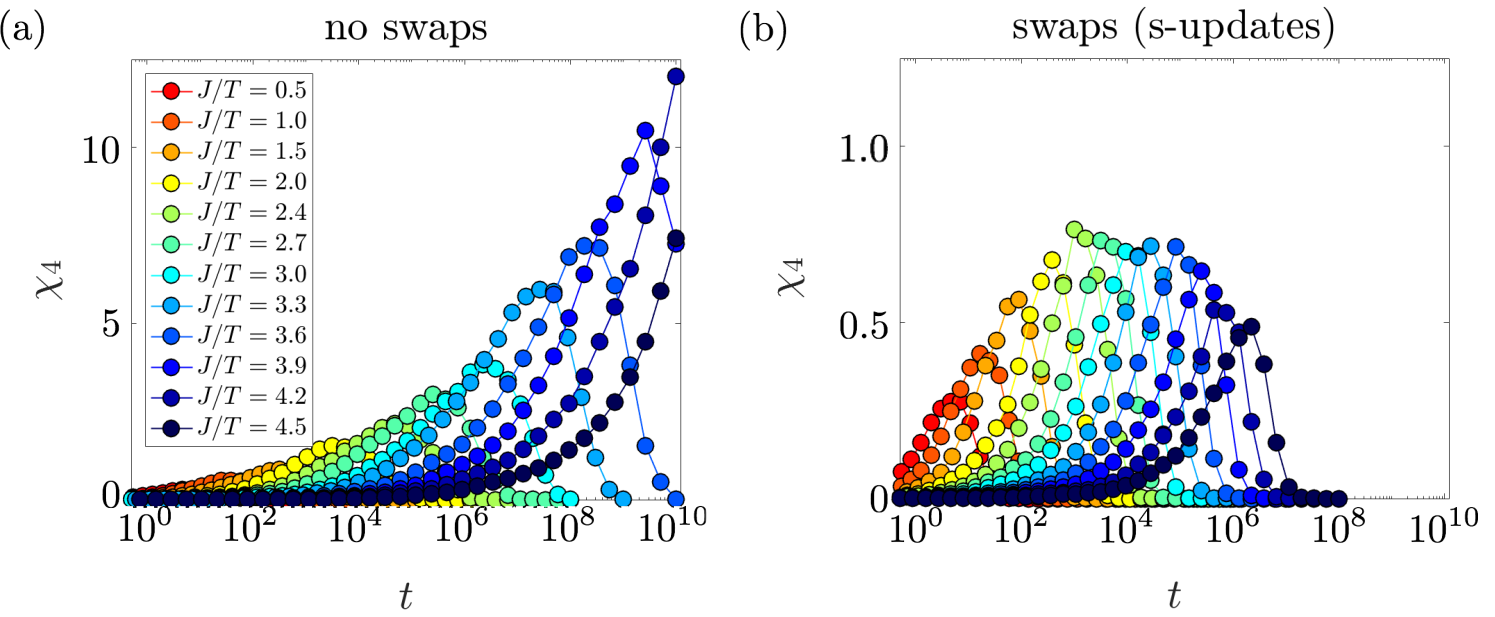}
\caption{Data for the dynamic susceptibility $\chi_4$ (see text for definition) without swaps (a) and with swaps (b) for different values of $J/T$ (see legend) with $B/v = 2J/T$ in a system of $N=512$ spins.  In the regime where the swaps lead to dramatic acceleration of the dynamics, the dynamics are not heterogeneous.
}
\label{fig:chi4}
\end{figure}

\subsection{Dynamical heterogeneity}

We consider the effect of swaps on the extent of heterogeneity in the dynamics. We quantify dynamical heterogeneity via the dynamic susceptibility
\beq
\chi_4(t) = \frac1N \left\langle \sum_{ij} [p_i(t) -P(t)] [p_j(t) - P(t)] \right\rangle.
\eeq
This function shows a maximum for some time $t^*\simeq\tau_{\rm p}$. If relaxation is dominated by the facilitated (East) mechanism then \rlj{the relaxation is heterogeneous with a characteristic length scale $\xi_4\sim \ee^{J/T}$ and one expects (in one dimension) that the maximal value of $\chi_4$ is proportional to this length scale, that is} $\chi_4(t^*) \sim \ee^{J/T}$ \cite{Berthier2005}.  If relaxation is dominated by a soft (non-collective) process then \rlj{relaxation much less heterogeneous, the length scale $\xi_4=O(1)$} and we expect $\chi_4(t^*)=O(1)$.  \rlj{Hence, the East relaxation process (which has a relaxation time growing faster than Arrhenius) also has a growing length scale and a growing $\chi_4$, as expected on general grounds.  On the other hand, the soft relaxation process has a small length scale and an Arrhenius growth of the relaxation time.}

Fig.~\ref{fig:chi4} shows $\chi_4(t)$ for different values of $J/T$ with $B/v = 2J/T$ (cf.\  Figs.~\ref{fig:pers} and \ref{fig:auto}). 
Results for no-swap dynamics are in Fig.~\ref{fig:chi4}(a), while in Fig.~\ref{fig:chi4}(b) shows the dynamics including swaps (s-updates). We observe the following: (i) the peak times $t^*$ are reduced in the presence of swaps, compatible with the results from the persistence and autocorrelation functions; (ii) dynamic heterogeneity is strongly suppressed in the presence of swaps (note that the vertical scale of both panels differs by an order of magnitude), because the soft process does not lead to heterogeneous relaxation; (iii) for the system with swaps, the peak height $\chi_4(t^*)$ depends non-monotonically on temperature, because the soft process becomes increasingly dominant at low temperature.

\begin{figure}
\begin{center}
\includegraphics[width=14cm]{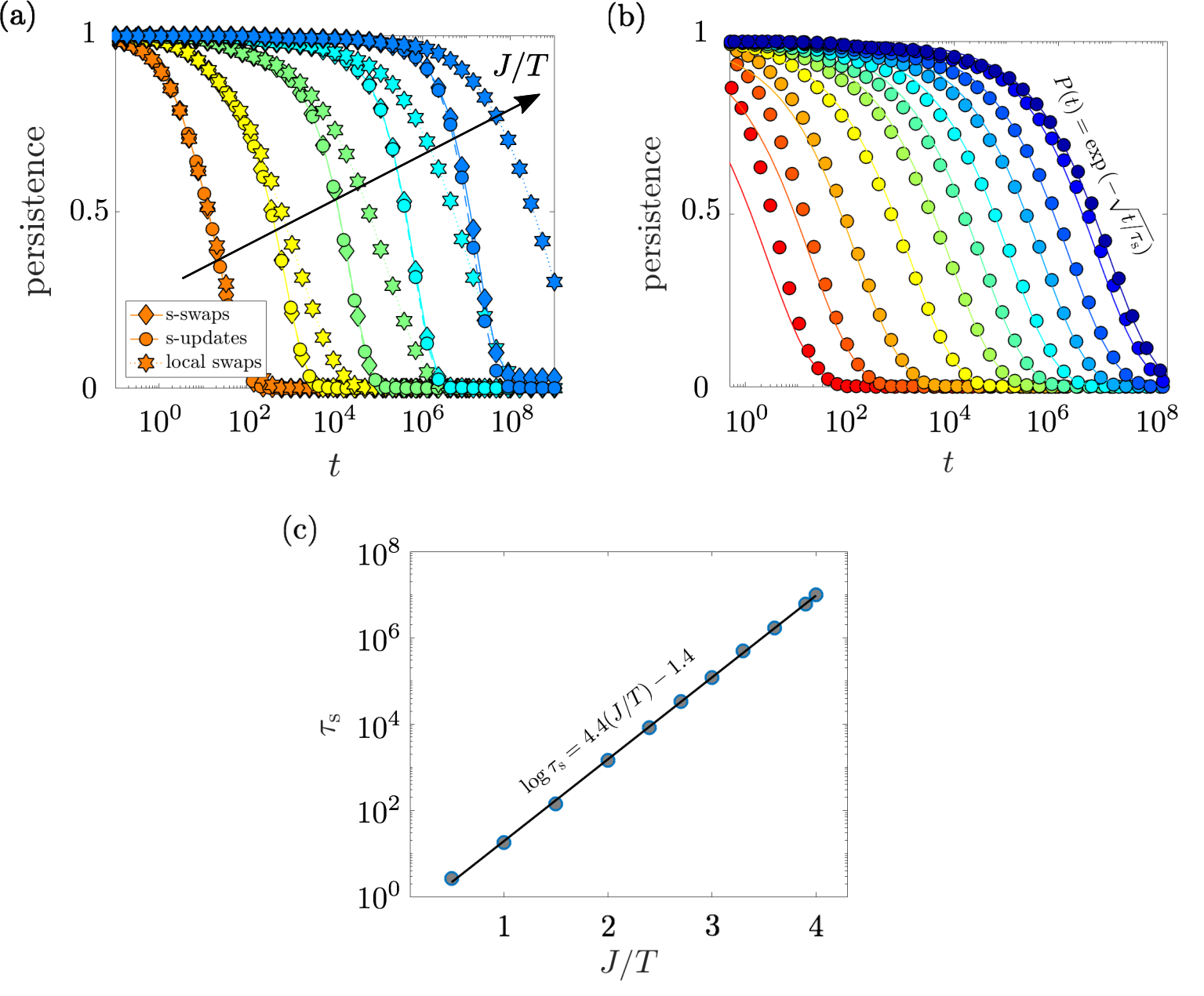}
\end{center}
\caption{(a) Persistence curves for different swap processes with $B/v = 2J/T$ in a system of $N=512$ spins. Different curves [corresponding to (inverse) temperatures $J/T = 1$, $2$, $3$, $4$ and $5$] are shown for (non-local) s-swaps, s-updates and local swaps. While s-swaps and s-updates give results that are essentially identical, as expected for sufficiently large system sizes, local swaps lead to a reduced accelerating effect on the relaxation dynamics. (b) Persistence curves for a dynamics with local swaps with $B/v=2J/T$ and $J/T = 0.5, 1, 1.5, 2, 2.4, 2.7, 3, 3.3, 3.6, 3.9, 4$ [some of these curves are displayed in panel (a) as well]. The solid lines are fits to $P(t) = \ee^{-\sqrt{t/\tau_{\rm s}}}$. (c) Characteristic time $\tau_{\rm s}$ [free parameter of the fit in panel (b)] as a function of $J/T$. A fit $\tau_{\rm s}\sim \ee^{bJ/T}$ shows excellent agreement with the numerical data. 
}
\label{fig:swap-upd}
\end{figure}

\subsection{Different kinds of swap process}
\label{subsec:swap-type}

Up to now we have considered swaps corresponding to s-updates, for both the real-valued and binary-softness models. 
Fig.~\ref{fig:swap-upd}(a) shows persistence curves for (non-local) s-swaps and for local swaps as well.  We take $r_{\rm s} = r_{\rm l}= r_{\rm u}/2 = \ee^{-J/T}/8$,  The value of $r_{\rm u}$ is consistent with previous sections.
The factor of two between $r_{\rm s}$ and $r_{\rm u}$ leads to similar behaviour for s-updates and s-swaps, because each s-swap is equivalent to an update of two spins.
The figure shows that s-swaps and s-updates behave almost identically, but local swaps lead to slower relaxation.  As argued in Subsection~\ref{subsubsec:swaps}, it is expected that s-updates and s-swaps should give almost identical results in large systems (as updating to a value chosen at random from the equilibrium distribution is equivalent to copying the value from a randomly-chosen site).   

It is also notable that the local swaps lead to non-exponential (stretched) relaxation at low temperatures, as seen in greater detail in Fig.~\ref{fig:swap-upd}(b). This behaviour is consistent with the discussion of Sec.~\ref{sec:local-swap-theory}.  The Figure shows a good fit to the stretched (non-exponential) relaxation predicted by (\ref{equ:P-fa}), with the relaxation time treated as a fitting parameter: that is $P(t) = \exp(-\sqrt{t/\tau_{\rm s}})$.  Using the scaling of $r_X,r_{\rm l}$ with temperature, (\ref{equ:pers-loc-swap}) predicts $\tau_{\rm s}\sim \ee^{bJ/T}$ with $b=5$: fits give $b=4.4$, see  Fig.~\ref{fig:swap-upd}(c) where $\tau_{\rm s}$ as a function of $J/T$ is shown.  This constitutes reasonable agreement, given that (\ref{equ:pers-loc-swap}) comes from a rather simplistic analogy with the FA model (recall the discussion in Sec.~\ref{sec:local-swap-theory}).   We note that this (significant) difference in relaxation time between systems with local and non-local swaps is not observed in atomistic models~\cite{berthier-local,Wyart2017}, as further discussed in Section~\ref{sec:discuss}.

\section{Discussion}
\label{sec:discuss}

It is an essential characteristic of KCMs that the relaxation time depends on the dynamical rules of the model, and cannot be determined by thermodynamic properties alone.  In this sense, the fact that the swap algorithm accelerates the structural relaxation of glass-formers~\cite{Grigera2001,Brumer2004,Fernandez2007,Gutierrez2015,Berthier2016,Ninarello2017} is entirely consistent with the dynamical facilitation theory of glasses~\cite{Chandler2010}. Our contribution here is to make this observation explicit by considering one specific mechanism by which swapping of local degrees of freedom can accelerate relaxation. In the KCM that we study, the relevant degree of freedom for the swap moves is the local softness, which can be interpreted as a local energy barrier associated with unfacilitated relaxation.

The mechanism of acceleration operates as follows: while the barriers at low temperature may be very high far away from excitations, in a particular region they can be temporarily reduced by swaps that increase the local softness (that is, by reducing the local barrier for the soft process which overcomes the constraint).  In the models that we have discussed, the swaps have a simple character, in that all sites have equal probability of participating in a swap move, and all such moves are accepted, independent of the local state.  The situation in atomistic models is more complicated, in particular, the ability of a region of the system to participate in a swap is likely to depend on its local structure.  Such an effect might help to explain some of the differences between the simple models presented here and atomistic liquids (see also below).  

This possibility might be incorporated in the approach shown here by modifying the acceptance probability of swap moves, so that it depends on the local X-values (see also~\cite{Bertin2005}).  As long as the dynamical rule always respects detailed balance with respect to the underlying equilibrium distribution, the theoretical arguments that we have given are easily generalised to this case, and we argue that the general picture that we present should be robust.  We also observe that the acceleration by swaps in our results is controlled by a small fraction of sites, which have $X_i>B$ (or $s_i=1$).  This aspect of the mechanism is natural within the framework considered here, but there might be other ways to combine KCMs with swap dynamics which would not have this property. 


Within the specific framework we considered here, the speedup by swaps is robust. Nevertheless, given the simplicity of our setting, there are some differences between the behaviour with atomisitic systems. We now comment on these. 

The acceleration of the dynamics by swaps in Fig.~\ref{fig:pers} is very significant: for the choice of parameters of the figure, the relaxation time becomes Arrhenius in the case with swaps in contrast to the super-Arrhenius behaviour of the original system. In contrast, in atomistic liquids it is found that swap dynamics, while faster, is still super-Arrhenius~\cite{Ninarello2017}.  In our case, the Arrhenius dependence is a consequence of the scaling of $B/v$, which we chose as proportional to $1/T$ in Fig.~\ref{fig:pers}.  It is not clear how the scaling of this quantity can be derived within dynamical facilitation theory, and a swap dynamics with super-Arrhenius temperature dependence could also be reproduced with soft KCMs by taking a different scaling of $B/v$ with $T$. However, this would require an explanation of why the energy barrier $B(T)$ should increase [a question that shares some similarities with the glass transition problem in general: how does $A(T)$ in (\ref{equ:tau-alpha}) change on cooling?].

A related point is that the non-monotonic dependence of $\chi_4^* = \chi_4(t^*)$ on temperature in Fig.~\ref{fig:chi4} is not seen in Fig. 14 of~\cite{Ninarello2017}, which shows analogous results for atomistic systems.  The residual growth of $\chi_4^*$ could naturally be associated with a residual collectiveness of the dynamics not completely removed by the swaps. It therefore seems plausible that a theory that could explain the temperature dependence of $B(T)$ would also explain the growth of $\chi_4*$ that occurs for dynamics with swaps. 

A final remark concerns the behavior of the model with local swap dynamics. In atomistic systems, local and non-local swaps lead to almost identical behaviour~\cite{berthier-local,Wyart2017}.  In Fig.~\ref{fig:swap-upd}, there is a significant difference in relaxation time between local and non-local swaps.  As discussed in Sec.~\ref{sec:local-swap-theory}, if both $r_{\rm l}$ and $r_{\rm s}$ are very large, one expects local and non-local swaps to lead to the same behaviour.  This limit is however difficult to study numerically as simulations become very inefficient.  In contrast, in atomistic simulations it is natural to choose larger swap rates than those considered here because the simulation of the atomistic dynamics is much more expensive than the swaps. A second consideration is that of dimensionality: Relaxation in these soft KCMs is mediated by rare anomalously soft sites, which move diffusively in the model with local swaps.  As discussed in Sec.~\ref{sec:local-swap-theory}, mixing by diffusion is slow in one dimension, compared with higher dimensions. It therefore seems likely that the differences between local and non-local swaps in Fig.~\ref{fig:swap-upd} would be much smaller in a corresponding three-dimensional version of the model.  (The change of dimensionality would not affect the hard constraint, as the features of the East model are mostly independent of spatial dimension~\cite{Berthier2005,Keys2011}.)

To conclude, acceleration by swaps can be reproduced quite naturally in kinetically constrained models.  Our work here provides a concrete and specific example of a model that reproduces this important phenomenon~\cite{Ninarello2017}. The acceleration due to swaps should be robust within KCMs, which makes the observation of enhanced relaxation in more realistic glass formers compatible with dynamic facilitation. In order to account more accurately for features such as the residual temperature dependence of timescales or the remnant growth of dynamical correlations (which in the context of KCMs are related to less coarse-grained features of supercooled liquids) further work is needed. This would be the natural direction in which to extend the models considered here.

\ack

RLJ thanks Ludovic Berthier for helpful discussions about swap dynamics. This work was funded by EPSRC Grants no.\ EP/M014266/1 and EP/R04421X/1.  We are also grateful for the computational resources and assistance provided by CRESCO, the super-computing center of ENEA in Portici, Italy.

\section*{Bibliography}

\bibliographystyle{iopart-num}
\bibliography{swaps,swaps-2}

\providecommand{\newblock}{}
\begin{thebibliography}{10}
\expandafter\ifx\csname url\endcsname\relax
  \def\url#1{{\tt #1}}\fi
\expandafter\ifx\csname urlprefix\endcsname\relax\def\urlprefix{URL }\fi
\providecommand{\eprint}[2][]{\url{#2}}

\bibitem{Binder2011}
Binder K and Kob W 2011 {\em Glassy materials and disordered solids: An
  introduction to their statistical mechanics\/} (World Scientific)

\bibitem{Berthier2011}
Berthier L and Biroli G 2011 {\em Rev. Mod. Phys.\/} {\bf 83} 587--645

\bibitem{Biroli2013}
Biroli G and Garrahan J~P 2013 {\em J. Chem. Phys.\/} {\bf 138} 12A301

\bibitem{Trachenko2011}
Trachenko K and Brazhkin V~V 2011 {\em Phys. Rev. B\/} {\bf 83}

\bibitem{Cavagna2009}
Cavagna A 2009 {\em Phys. Rep.\/} {\bf 476} 51--124

\bibitem{Chandler2010}
Chandler D and Garrahan J~P 2010 {\em Annu. Rev. Phys. Chem.\/} {\bf 61}
  191--217

\bibitem{Berthier2007mc}
Berthier L and Kob W 2007 {\em J. Phys.: Cond. Matt.\/} {\bf 19} 205130

\bibitem{Grigera2001}
Grigera T~S and Parisi G 2001 {\em Phys. Rev. E\/} {\bf 63} 045102

\bibitem{Brumer2004}
Brumer Y and Reichman D~R 2004 {\em J. Phys. Chem. B\/} {\bf 108} 6832--6837

\bibitem{Fernandez2007}
Fernandez L~A, Martin-Mayor V and Verrocchio P 2007 {\em Philos. Mag.\/} {\bf
  87} 581--586

\bibitem{Gutierrez2015}
Guti{\'e}rrez R, Karmakar S, Pollack Y~G and Procaccia I 2015 {\em EPL\/} {\bf
  111} 56009

\bibitem{Berthier2016}
Berthier L, Coslovich D, Ninarello A and Ozawa M 2016 {\em Phys. Rev. Lett.\/}
  {\bf 116} 238002

\bibitem{Ninarello2017}
Ninarello A, Berthier L and Coslovich D 2017 {\em Phys. Rev. X\/} {\bf 7}
  021039

\bibitem{Berthier2017entropy}
Berthier L, Charbonneau P, Coslovich D, Ninarello A, Ozawa M and Yaida S 2017
  {\em Proc. Natl. Acad. Sci. USA\/} {\bf 114} 11356--11361

\bibitem{Ozawa2018}
Ozawa M, Berthier L, Biroli G, Rosso A and Tarjus G 2018 {\em Proc. Natl. Acad.
  Sci. USA\/} {\bf 115} 6656--6661

\bibitem{Wang2019}
Wang L, Ninarello A, Guan P, Berthier L, Szamel G and Flenner E 2019 {\em
  Nature Communications\/} {\bf 10} 26

\bibitem{Wyart2017}
Wyart M and Cates M~E 2017 {\em Phys. Rev. Lett.\/} {\bf 119} 195501

\bibitem{Brito2018}
Brito C, Lerner E and Wyart M 2018 {\em Phys. Rev. X\/} {\bf 8} 031050

\bibitem{Berthier2019-jcp}
Berthier L, Biroli G, Bouchaud J~P and Tarjus G 2019 {\em J. Chem. Phys.\/}
  {\bf 150} 094501

\bibitem{Ikeda2017}
Ikeda H, Zamponi F and Ikeda A 2017 {\em The Journal of Chemical Physics\/}
  {\bf 147} 234506

\bibitem{Szamel2018}
Szamel G 2018 {\em Phys. Rev. E\/} {\bf 98} 050601

\bibitem{Ritort2003}
Ritort F and Sollich P 2003 {\em Adv. Phys.\/} {\bf 52} 219--342

\bibitem{Garrahan2011}
Garrahan J~P, Sollich P and Toninelli C 2011 {Kinetically Constrained Models}
  {\em Dynamical Heterogeneities in Glasses, Colloids, and Granular Media\/}
  International Series of Monographs on Physics ed Berthier L, Biroli G,
  Bouchaud J~P, Cipelletti L and van Saarloos W (Oxford, UK: Oxford University
  Press)

\bibitem{Pan2005}
Pan A, Garrahan J and Chandler D 2005 {\em {ChemPhysChem}\/} {\bf 6} 1783--1785

\bibitem{Elmatad2010}
Elmatad Y~S, Jack R~L, Chandler D and Garrahan J~P 2010 {\em Proc. Natl. Acad.
  Sci. USA\/} {\bf 107} 12793--12798

\bibitem{Shokef2010}
Shokef Y and Liu A~J 2010 {\em {EPL} (Europhysics Letters)\/} {\bf 90} 26005

\bibitem{Keys2011}
Keys A~S, Hedges L~O, Garrahan J~P, Glotzer S~C and Chandler D 2011 {\em Phys.
  Rev. X\/} {\bf 1} 021013

\bibitem{Jackle1991}
J\"{a}ckle J and Eisinger S 1991 {\em Z. fur Phys. B\/} {\bf 84} 115--124

\bibitem{Elmatad2013}
Elmatad Y~S and Jack R~L 2013 {\em J. Chem. Phys.\/} {\bf 138} 12A531

\bibitem{Gutierrez2016}
Guti{\'e}rrez R and Garrahan J~P 2016 {\em J. Stat. Mech.\/} {\bf 2016} 074005

\bibitem{Sollich1999}
Sollich P and Evans M~R 1999 {\em Phys. Rev. Lett.\/} {\bf 83} 3238

\bibitem{Chleboun2013}
Chleboun P, Faggionato A and Martinelli F 2013 {\em J. Stat. Mech.\/} {\bf
  2013} L04001

\bibitem{Berthier-new-swap}
Berthier L, Flenner E, Fullerton C~J, Scalliet C and Singh M arXiv:1811.12837

\bibitem{Garrahan2002}
Garrahan J~P and Chandler D 2002 {\em Phys. Rev. Lett.\/} {\bf 89} 035704

\bibitem{Jung2005}
Jung Y, Garrahan J~P and Chandler D 2005 {\em J. Chem. Phys.\/} {\bf 123}
  084509

\bibitem{Pan2005-het}
Pan A, Garrahan J and Chandler D 2005 {\em Phys. Rev. E\/} {\bf 72} 041106

\bibitem{Teomy2015}
Teomy E and Shokef Y 2015 {\em Phys. Rev. E\/} {\bf 92}(3) 032133

\bibitem{Bortz1975}
Bortz A~B, Kalos M~H and Lebowitz J~L 1975 {\em J. Comput. Phys.\/} {\bf 17}
  10--18

\bibitem{Berthier2005}
Berthier L and Garrahan J~P 2005 {\em J. Phys. Chem. B\/} {\bf 109} 3578--3585

\bibitem{berthier-local}
L. Berthier, private communication

\bibitem{Bertin2005}
Bertin E, Bouchaud J and Lequeux F 2005 {\em Phys. Rev. Lett.\/} {\bf 95}
  015702

\end{thebibliography}

\end{document}